\documentstyle[12pt]{article}
\newcommand{\k}{{\bf\hat K}}
\newcommand{\s}{{\bf\hat S}}
\newcommand{\bm}{\mbox{B}(m)}
\newcommand{\bg}{\hat{\mbox{B}}(\hat G_z)}

\input epsf

\begin{document}

\begin{center}
{\Large\bf Extracting Hidden Symmetry from the Energy Spectrum}
\vspace{1cm}

Emil~A.~Yuzbashyan$^1$, William Happer$^1$,
Boris~L.~Altshuler$^{1,2}$, Sriram~B.~Shastry$^3$

\vspace{0.5cm}

{\small\it

\noindent $^1$ Physics Department, Princeton University, Princeton, NJ 08544\\
 $^2$ NEC Research Institute, 4 Independence Way,
Princeton, NJ 08540\\
$^3$  Physics Department, Indian Institute of Science, Bangalore
560012, India}
\end{center}

\begin{abstract}

In this paper we revisit the problem of finding hidden symmetries
in  quantum mechanical systems. Our interest in this problem was
renewed by nontrivial degeneracies of a simple spin Hamiltonian used
to model spin relaxation in alkali-metal vapors. We consider this
spin Hamiltonian in detail and use this example
to outline a general approach to
 finding symmetries when eigenvalues and
eigenstates of the Hamiltonian are known. We extract all
nontrivial symmetries responsible for the degeneracy and show that
the symmetry group of the Hamiltonian is $SU(2)$. The symmetry
operators have a simple meaning which becomes transparent in the
limit of large spin. As an additional example we apply the method
to the Hydrogen atom.

\end{abstract}

\section{Introduction}

The close connection between symmetry and degeneracy has been explored since
the very foundation of Quantum Mechanics.
 Famous examples include  degeneracies
of the energy
 spectra in angular momentum in the 3d harmonic oscillator \cite{LL}
and in the Hydrogen
atom (well known as the accidental degeneracy) \cite{Pauli}.
Here we undertake a
detailed study of the connection between symmetries  and  degeneracies of
a Hamiltonian that describes the exchange interaction of two spins,
$\s$ and $\k$, and also includes Zeeman splitting for spin $\s$:
\begin{equation}
\hat H(x)=x(K+1/2) \hat S_z+\k\cdot\s
\label{Hx}
\end{equation}
where $S=1$ and $K$ is arbitrary.

\begin{figure}[ht]
\epsfxsize 15 cm \centerline{\epsfbox{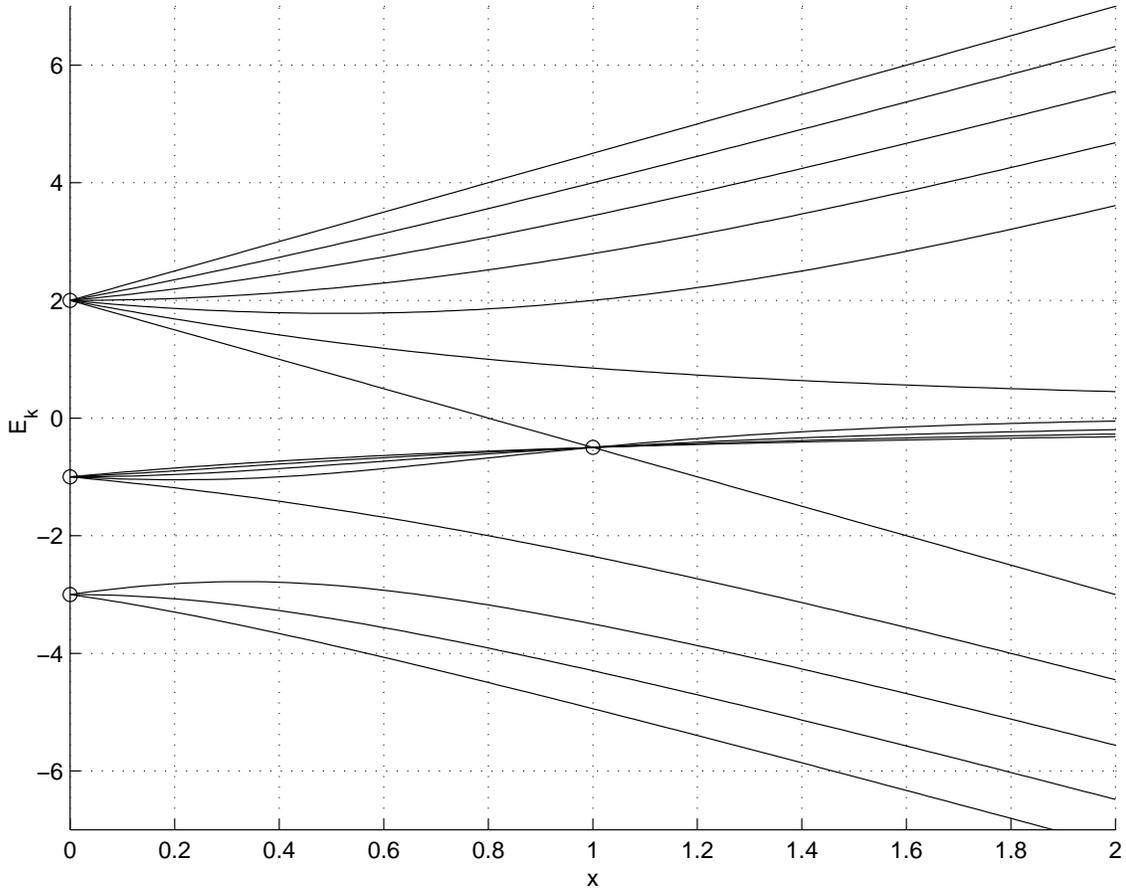}}
\caption{Eigenvalues of $\hat H(x)=x(K+1/2) \hat S_z+\k\cdot\s$ for $S=1$ and
$K=2$. Note the level crossings at $x=1$.} \label{levels}
\end{figure}

Our
interest in this system was motivated by experiments
\cite{Will} on spin relaxation in  polarized alkali-metal vapors.
In this case $\s$ has a meaning of the total electronic spin, $\k$
is the nuclear spin, and the dimensionless constant $x$ represents
the magnetic field.  A diagram of energy levels of
 Hamiltonian (\ref{Hx}) for a typical value of $K=2$ is shown on
Fig.~\ref{levels}.
 In addition to
 degeneracies at $x=0$ (no Zeeman splitting) the spectrum displays
 less trivial $(2K+1)$-fold degeneracies at $x=\pm1$. These degeneracies
 show up
as resonances in the spin relaxation rate and provide the key evidence for
a particular mechanism of spin relaxation (see \cite{Will} for
details). The degeneracies at $x=\pm1$ have been also discussed in \cite{guys}.

Despite the simplicity of  Hamiltonian (\ref{Hx}), it is not
trivial to determine  its symmetry at $x=1$. To the best of
our knowledge  there is no  general textbook algorithm
for extracting symmetries. In this paper we propose such an
algorithm. We demonstrate how,  using only a minimal intuition into the
reasons for degeneracy, one
can
 find hidden symmetries
whenever the eigenstates and the eigenvalues of the Hamiltonian are
known.

Before proceeding further let us  formulate the general problem of
identifying the symmetry responsible for a particular spectral
degeneracy. Symmetries manifest themselves through operators that
commute with the Hamiltonian, $\hat H$. Consider a set of these operators
${\cal A}=\{\hat A_i\}$. If not all of the operators $\hat A_i$
commute with each other, this symmetry implies some spectral
degeneracy. Indeed, if
\begin{equation}
[\hat A_i, \hat H]=[\hat A_j, \hat H]=0\quad\mbox{and}\quad
[\hat A_i, \hat A_j]\ne0,
\label{triv}
\end{equation}
there exists an eigenstate of the Hamiltonian $|\alpha\rangle$ for
which $\hat A_i|\alpha\rangle\ne \hat A_j|\alpha\rangle$. This
state is necessarily degenerate, since according to equation
(\ref{triv}) $\hat A_i|\alpha\rangle$ and $\hat A_j|\alpha\rangle$
are eigenstates of $\hat H$ of the same energy as
$|\alpha\rangle$ (see e.g. \cite{LL}).

Is a particular set of operators ${\cal A}$ sufficient to account
for given spectral degeneracies? The positive answer to this
question implies that
\begin{enumerate}
\item all degenerate states can be labelled by eigenvalues of a number
  of mutually commuting operators from the set ${\cal A}$
\item any degenerate state can be obtained from any other state in the
  same eigenspace by a repeated action of operators $\hat A_i$ or their
  linear combinations.
\end{enumerate}

Provided that conditions 1) and 2) are met one can complete the
analysis by clarifying the physical significance of operators $\{\hat A_i\}$.
An additional, more abstract,
 question one might ask is what is the group
generated by operators $\{ \hat A_i\}$.

Consider for example the degeneracy of Hamiltonian (\ref{Hx})
at $x=0$. One can introduce ${\bf \hat G}=\k+\s$ and
identify, e.g., $\hat A_1=\hat G_x$ and $\hat A_2=\hat G_y$.
The commutator of $\hat A_1$ with $\hat A_2$ yields $\hat A_3=\hat G_z$.
Eigenvalues of $\hat G_z$ and ${\bf\hat G}^2$ can be used to label the
degenerate states, while linear combinations of $\hat G_x$ and $\hat G_y$,
$\hat G_+$ and $\hat G_-$, connect all degenerate states in a given
eigenspace. Finally, any  operator that commutes with $\hat H(0)$ can
be written in terms of $\hat G_x$, $\hat G_y$, $\hat G_z$, and
${\bf\hat G}^2$. The
group generated by the components of ${\bf\hat G}$ is $SU(2)$ and the
group elements have a meaning of rotations in 3d space.

Frequently, we encounter a situation when the spectrum is known
exactly and yet the symmetries are hidden. In such cases the
following formal expression for commuting operators serves as a
useful starting point. One can show quite generally that, up to an
operator that annihilates all degenerate states,
all integrals of motion must be of the
form
\begin{equation}
\hat X=\sum_n \hat P_n \hat Y \hat P_n\label{main}
\end{equation}
Here $\hat P_n$ is an operator that projects out the $n$th degenerate
eigenspace and $\hat Y$ is an arbitrary operator. We prove  equation
(\ref{main})
 in  the Appendix.  The problem of
finding the symmetry thus reduces to making appropriate choices for
$\hat Y$. This choice can be either guided by limiting cases, e.g., the
classical limit, where the symmetry is simple to identify, or by an
 intuition as to what type of conservation laws (e.g.
scalars or vectors or etc.) one expects to find. However,
 examples we considered show that
many apparently different choices for the operator $\hat Y$ produce
equivalent
 conservation
laws. Thus, it is usually sufficient to explore the simplest
options - the basic operators of the problem. For example, if the
Hamiltonian is written in terms of ${\bf\hat r}$ and ${\bf\hat p}$,
natural choices for $\hat Y$ would be ${\bf\hat r}^2$ and ${\bf\hat p}^2$ if
one is looking for a conserved scalar or ${\bf\hat r}$ and ${\bf\hat p}$
if a conserved vector is expected.

The paper is organized as follows. First, we determine and analyze
in detail the spectrum of Hamiltonian (\ref{Hx}) at $x=1$ (Section
2). We show that, apart from the zero field degeneracy at
$x=0$, there are degeneracies only at $x=\pm1$, the spectrum at
$x=-1$ being simply related to that at $x=1$.  Given the
eigenstates, integrals of motion (\ref{main}) can be
evaluated without knowing projectors $\hat P_n$ explicitly. In this
case the general strategy is to compute the matrix elements of $\hat X$
in equation (\ref{main}) and use them to determine $\hat X$ in terms of
basic operators of the problem. We will illustrate this approach
in
Section 6.

In Section 3 we use (\ref{main}) to explicitly derive commuting operators that
connect all degenerate states at $x=1$. We demonstrate that  the
symmetry group of Hamiltonian (\ref{Hx}) at $x=1$ is $SU(2)$. The
physical meaning of symmetry operators is clarified in Section 4,
where we consider the limit of large $K$.

In
Section 5 we rewrite the Hamiltonian in the form that makes the
analogy to the large $K$ limit particularly clear and allows us to
establish some additional properties of the energy spectrum. In
Section 6 we provide another example of the general
approach of finding symmetries based on (\ref{main}) by deriving
the Runge-Lenz vector from (\ref{main}).

\section{Energy Spectrum}

Here we analyze the spectrum of Hamiltonian (\ref{Hx}) at arbitrary
$x$ and consider in detail the spectrum at $x=1$.
The results  discussed
in this section were originally derived  in \cite{Will2}.
We will follow \cite{Will2} closely adopting in most cases the same notation.

 First, we note that the $z$ component of the
total spin is conserved.
\begin{equation}
[\hat G_z, \hat H(x)]=0
\label{Gz}
\end{equation}
Since $S=1$,  there are at most three
independent states for each eigenvalue $m$ of $\hat G_z$.
In the basis $|K_z,  S_z \rangle$ these states are
$|m-1, 1\rangle$, $|m, 0\rangle$, and $|m+1, -1\rangle$. The
block of the Hamiltonian (\ref{Hx}) that corresponds  to $m$
for $|m|\le K-1$ is
\begin{equation}
\hat H=\frac{1}{\sqrt{2}}\left[\begin{array}{ccc}
\sqrt{2}(x(K+1/2)+m-1)& \sqrt{(K+m)(K-m+1)}& 0 \\
\sqrt{(K+m)(K-m+1) }& 0& \sqrt{(K-m)(K+m+1)}\\
0&  \sqrt{(K-m)(K+m+1)}& \sqrt{2}(-x(K+1/2)-m-1)\\
\end{array}\right]
\label{block}
\end{equation}
with the eigenvalue equation
\begin{equation}
\begin{array}{l}
 E^3+2E^2-a(x)E-K^2-b(x)=0\\
\\
a(x)=(1+x^2)(K+1/2)^2+2x(K+1/2) m-5/4\\
\\
b(x)=K^2+K+x(K+1/2)m\\
\end{array}
\label{EE}
\end{equation}
In addition to $3\times3$ blocks
(\ref{block})
the Hamiltonian also has two $2\times2$
blocks for $|m|=K$ and two $1\times1$ blocks for $|m|=K+1$. Energies
 for these values of $m$ are also solutions of (\ref{EE}).

Degeneracies occur only at $x=0$ and $x=\pm1$ (see
Fig.~\ref{levels}). The energy spectrum at $x=-1$ is identical to that at $x=1$
while the eigenstates are related via a unitary
transformation (a rotation by $\pi$ around any axis in $xy$
plane). Therefore, it is sufficient to consider only $x=1$.

 The spectrum at $x=1$ has the following
features:
\begin{itemize}
\item There are $2K+1$ degenerate states $|Dm\rangle$
(``D'' for degenerate states) with the energies ${E_{Dm}=-1/2}$

\item There is a gap at $m=-K$ in the values of $m$ that can
be assigned
to degenerate states  $|Dm\rangle$, $m=K, K-1,\dots,-K+1$ and $-K-1$
\item There are $2K+2$ non-degenerate states $|Tm\rangle$ (``T'' for top
states) with energies $E_{Tm}>-1/2$ and $m=K+1, K,\dots,-K$
\item There are $2K$ non-degenerate states $|Bm\rangle$ (``B'' for bottom
states) with energies $E_{Bm}<-1/2$ and $m=K-1, K,\dots,-K$
\item Energies of the top $|Tm\rangle$ and bottom  $|Bm\rangle$ states
are related by $E_{Tm}+E_{Bm}=-3/2$ ($|m|<K$)
\end{itemize}

Finally, we can use (\ref{block}) to compute the energies at $x=1$ and
the wave-functions $|Dm\rangle$ of the
degenerate states
\begin{equation}
E_{Dm}=-1/2\quad E_{Tm}=-3/4+\sqrt{\bm^2+1/16}
\quad E_{Bm}=-3/2-E_{Tm}
\label{energies}
\end{equation}
\begin{equation}
\left[ \begin{array}{l}
\langle m-1,1|Dm\rangle\\
\langle m,0|Dm\rangle\\
\langle m+1,-1|Dm\rangle\\
\end{array}\right]=
 \frac{1}{\sqrt{2}\bm}\left[\begin{array}{c}
-\sqrt{(K+1+m)(K+1-m)}\\
\sqrt{2(K+1+m)(K+m)}\\
\sqrt{(K+m)(K-m)}\\
\end{array}\right]
\label{wavefn}
\end{equation}
where
\begin{equation}
\bm=(2K+1)(K+m+1/2)
\label{b}
\end{equation}

\section{Symmetries}

In this section we derive integrals of motion responsible for
degeneracies at $x=1$. We write them in terms of components of $\s$ and
$\k$ and show that the symmetry group is $SU(2)$.

Our starting point is the general expression for commuting operators
(\ref{main}). Hamiltonian (\ref{Hx}) has only one degenerate subspace
at $x=1$, so there is only one projection operator in (\ref{main}).
\begin{equation}
\hat X=\hat P\hat Y\hat P
\label{commain}
\end{equation}
 Even though we do not need to know projection operators explicitly to
 evaluate (\ref{main}), in the case of Hamiltonian (\ref{Hx}) the
 projector
 $\hat P$ has a
simple meaning in the large $K$ limit, so we derive it bellow
from the eigenvalue equation (\ref{EE}).

 When the energy $E$ is
replaced by $\hat H$ and $m$ is replaced by $\hat G_z$,  eigenvalue
equation
(\ref{EE})
translates into a cubic identity for the Hamiltonian \cite{Will2}.
At $x=1$ this identity can be factored out as
follows
\begin{equation}
(\hat H+1/2)\underbrace{\left[\bg^2-
(\hat H+1/2)(\hat H+1)\right]}_{\hat \Pi}=0
\label{idfactored}
\end{equation}
where
\begin{equation}
\bg=(2K+1)(K+\hat G_z+1/2)
\label{bg}
\end{equation}

States with $E=-1/2$ are degenerate with respect to the eigenvalues of
$\hat G_z$. The operator in the
square brackets, $\hat \Pi$, is then an ``unnormalized projection
operator''
that
projects out the degenerate subspace.
\begin{equation}
\hat \Pi|Nm\rangle=0\quad \hat \Pi|Dm\rangle=\bm^2|Dm\rangle
\label{Ponst}
\end{equation}

The operator $\hat \Pi$ can be normalized
to a usual projection operator for the degenerate
subspace \cite{Will2}.
\begin{equation}
\hat P=\frac{\hat \Pi}{\bg^2}=
1-\frac{(\hat H+1/2)(\hat H+1)}{\bg^2}
\label{Pn}
\end{equation}

Now we have to identify appropriate choices for the operator  $\hat Y$
in equation (\ref{commain}). Let us first try the simplest options
- operators that are linear in components of $\s$ and $\k$.
Consider, for example, a conserved vector\footnote{Projection
operator $\hat P$ is
  a scalar, because the Hamiltonian is a scalar. To see this
 the operator $\hat S_z$ in Hamiltonian (\ref{Hx}) should be written
  as $\s\cdot{\bf b}$, where ${\bf b}$ is a unit vector along the
  magnetic field. By the same argument $\hat G_z={\bf\hat G\cdot b}$ is
a conserved scalar. Therefore, ${\bf L}$ is a vector.}
\begin{equation}
{\bf \hat L}=\hat P\k\hat P
\label{L}
\end{equation}
Let us determine the action of components of ${\bf \hat L}$ on the
eigenstates of  Hamiltonian. Since $\hat P$ is a projector for the
degenerate subspace\footnote{We use the usual notation
$\hat L_1\equiv \hat L_x\equiv\frac{\hat L_++\hat L_-}{2}$, $\hat L_2\equiv
\hat L_y\equiv
\frac{\hat L_+-\hat L_-}{2i}$, and $\hat L_3\equiv\hat L_z$.} ,
$$
\hat L_i|Nm\rangle=0\qquad i=1, 2, 3.
$$
To calculate the action of $\hat L_i$ on degenerate states
recall that
due to selection rules (see e.g. \cite{LL}) the
components of a vector can have nonzero matrix elements only for
transitions $m\to m$ and $m\to m\pm1$. We note also the following
relation between matrix elements of operators $\hat X$ and $\hat Y$ in
equation
(\ref{commain})
\begin{equation}
\langle m'|\hat X|m\rangle=\langle Dm'|\hat Y|Dm\rangle
\label{mel}
\end{equation}
where $|m\rangle$ and $|m'\rangle$ are
 any two eigenstates of $\hat G_z$ with eigenvalues $m$ and $m'$
 respectively. Using equations (\ref{mel}) and (\ref{wavefn}), we obtain
\begin{equation}
\hat L_+|Dm\rangle=
\sqrt{\frac{(K^2-m^2)((2K+2m+3)^2-1)}{(2K+2m+2)^2-1}}
|D,m+1\rangle
\label{plus}
\end{equation}
The action of $\hat L_-$ can be determined directly from equation
(\ref{plus})
 and
the action of $\hat L_z$ from equations (\ref{mel}) and
(\ref{wavefn})
with $m'=m$.

Note that
equation (\ref{plus}) is not well
 defined for $m=-K-1$. For the state $|D, -K-1\rangle$
instead of
equation (\ref{plus}) we get
\begin{equation}
\hat L_\pm|D, -K-1\rangle=0
\label{excl}
\end{equation}
Equation (\ref{excl}) reflects the existence of a gap at $m=-K$ in
the eigenvalues of $\hat G_z$ assigned to degenerate states. An
operator that connects the state $|D, -K-1\rangle$ to the rest of
degenerate subspace must change $m$ by at
least two. According to  selection rules such an operator
is neither a scalar nor a component of a vector.  Let
us  first consider values of $m\ne -K-1$ and revisit the problem of
the gap at the end of the section.

Since equation (\ref{plus}) is not of
the standard form for a raising operator in $su(2)$, commutation
relations for $\hat L_\pm$ are ``deformed'' versions of $su(2)$
commutation relations. However, one can derive usual $su(2)$
operators from $\hat L_\pm$ and $\hat G_z$. Define
\begin{equation}
\hat N_+=(\hat N_-)^\dagger=\hat A(\hat G_z) \hat A(\hat G_z+1) \hat L_+\qquad
\hat N_z=(\hat G_z-1/2)\hat P
\label{smallsu2}
\end{equation}
where
$$
\hat A(\hat G_z)=\frac{\bg}{\hat{\mbox{B}}(\hat G_z+1/2)}
$$
Operators $\hat N_i$ form an $su(2)$ algebra:
$$
[\hat N_i, \hat N_j]=\varepsilon_{ijk}\hat N_k
$$
All degenerate states with $m\ne
-K-1$ are connected by operators $N_\pm$.
 Each of these states is labelled by  $N_z$ and by an
eigenvalue of the Casimir operator
$$
{\bf\hat N}^2\equiv
\sum_{i=1}^{3} \hat N_i^2=\hat N_-\hat N_++\hat N_z^2+\hat N_z
$$
Operator ${\bf\hat N}^2$ has eigenvalues $N(N+1)$ with (half)integer
values of $N$.
$$
|Dm\rangle=|\: N=K-1/2,  N_z=m-1/2\rangle\quad m\ne -K-1
$$
Thus, degenerate states with $m\ne -K-1$ transform under the
$2K$-dimensional representation of this $su(2)$, while
nondegenerate states are singlets.

Note that by selection rules taking any other  vector, e.g. a
linear combination of $\k$ and $\s$, instead
of
$\k$ in the definition
of ${\bf \hat L}$, equation (\ref{L}), will produce the same results and
lead to the same $su(2)$ algebra.

As was mentioned above,
to connect the state $|D,
-K-1\rangle$ to the rest of the degenerate subspace, one needs an
operator that can change $m$ by at least two. Let us consider, for
example, operators
\begin{equation}
\hat M_\pm= \hat P \hat S_\pm^2 \hat P
\label{M}
\end{equation}
Using operators $\hat M_\pm$ and operators $\hat N_i$, one can construct a new
$su(2)$ algebra with a representation that incorporates the state
$|D, -K-1\rangle$. We define
\begin{equation}
\begin{array}{l}
\hat J_+=(\hat J_-)^\dagger=\sqrt{\frac{K+\hat G_z}{K+\hat
    G_z-1}}\hat N_+ -\sqrt{3K+3/2}\:\;
\hat P\hat S_+^2\hat P_{-K-1}\\
\\
 \hat J_z=\hat G_z\hat P+\hat P_{-K-1}\\
\end{array}
\label{bigsu2}
\end{equation}
Here\footnote{Note that $\hat P_{-K-1}=\hat P_{-K-1}\hat P=
\hat P\hat P_{-K-1}\hat P$ and
$\hat G_z\hat P=\hat P\hat G_z=\hat P\hat G_z\hat P$.
Hence, operators $J_\pm$ and $J_z$ defined
by equations (\ref{bigsu2}) are indeed of the general form (\ref{main}).}
$\hat P_{-K-1}$ denotes the projector onto the state $|D,-K-1\rangle$
$$
\hat P_{-K-1}|D,-K-1\rangle=|D,-K-1\rangle\qquad
\hat P_{-K-1}|m\ne-K-1\rangle=0
$$
The operator $\hat P_{-K-1}$ can be written in terms of $\hat G_z$ as follows
$$
\hat P_{-K-1}=\frac{1}{(2K+2)!}\prod_{m\ne -K-1} (m-\hat G_z)
$$

One can check by a direct computation using equations
(\ref{bigsu2}), (\ref{smallsu2}), (\ref{plus}), and (\ref{Pn}) that
\begin{enumerate}
\item Operators $\hat J_\pm$ and $\hat J_z$ commute with Hamiltonian
(\ref{Hx}) at
$x=1$ and form an $su(2)$ algebra.

\item All degenerated states are uniquely labelled by eigenvalues of
  $\hat J_z$ and ${\bf\hat J}^2$
$$
|Dm\rangle=|J=K,  J_z=m\rangle\quad -K+1\le m\le K
$$
and
$$
|D, -K-1\rangle=|J=K,  J_z=-K\rangle
$$
Thus, $J_z$ runs from $-K$ to $K$ with no gap and the degenerate subspace
transforms under a (2K+1)-dimensional representation of the $su(2)$
defined by equations (\ref{bigsu2}).

\item All nondegenerate states are singlets, i.e. they have $J=J_z=0$
\end{enumerate}

Note that
operators $\hat M_\pm$ are linear combinations of components of a conserved
rank 2 tensor $\hat M_{ij}=\hat P\hat S_i\hat S_j\hat P$. By the same
argument
 as for vectors,
inserting
any other rank 2 tensor instead of operator $\hat Y$ in equation
(\ref{commain}) will result in equivalent operators. Higher rank tensors
are not needed. Indeed, operators $\hat J_i$ already
connect all
degenerate states. Therefore, components of tensors of a rank
greater than two can be written in terms of products of operators
$\hat J_i$ in
the same way as components of any tensor that commutes with
Hamiltonian (\ref{Hx}) at $x=0$ can be expressed through $\hat G_x$,
$\hat G_y$,
and $\hat G_z$.

\section{Large $K$ Limit}

To gain some insight into the conservation laws derived in the previous
section
 let us
consider the limit $K\gg 1$.
In this limit the meaning of symmetry turns out to be transparent. Our
 intention in
this section is not to present a rigorous analysis, but rather to
develop an intuition about symmetries responsible for degeneracy.

 If $\k$ is a classical vector, we can write the Hamiltonian at $x=1$ as
\begin{equation}
\hat H_{cl}=|(K+1/2) {\bf z}+\k|\hat S_{\bf n}\label{Hcl}
\end{equation}
where ${\bf z}$ is a unit vector along the $z$-axis and operator
$\hat S_{\bf n}$ is the
projection of $\s$ onto the axis parallel to the vector $(K+1/2){\bf z}+\k$.
For each orientation of $\k$
there are three
energies corresponding to $S_{\bf n}=+1,$ $-1$ and $0$ -- ``top'',
``bottom'' and ``middle'' levels .
 Middle levels have zero energy, independent on the direction of $\k$.

 Now let us take into account quantization of $\k$ in the limit
$K\gg 1$. We assume  nevertheless that $\hat S_{\bf n}$
still can be interpreted as a projection of $\s$
\footnote{The large $K$ limit can be treated more accurately
  by introducing a preferred direction for $\k$ and representing
$\k$ in terms of Holstein--Primakoff bosons.}.

The direction of $\k$ is specified by $K_z\approx m$. States
${|S_{\bf n}=0,  G_z=m\rangle}$ are degenerate with respect to $m$.
 Evidently, symmetries responsible for this degeneracy
``rotate $\k$'' for states with $\hat S_{\bf n}=0$,
while keeping it unchanged
for states with $\hat S_{\bf n}\ne0$. Generators for such
rotations are
\begin{equation}
\hat L^{cl}_{x, y}=(1-\hat S_{\bf n}^2)\hat K_{x, y}(1-\hat S_{\bf n}^2)
\label{Lcl}
\end{equation}
Operators $\hat L^{cl}_{x, y}$ commute
with the Hamiltonian (\ref{Hcl}) which follows from the spin-1
 identity
\begin{equation}
\hat S_{\bf n}(1-\hat S_{\bf n}^2)=0
\label{spiden}
\end{equation}
In deriving (\ref{Lcl}) we made use of a projection operator
\begin{equation}
\hat P_{cl}=1-\hat S_{\bf n}^2\qquad
\hat P_{cl}|Dm\rangle=|Dm\rangle\quad \hat P_{cl}|Nm\rangle=0
\label{Pcl}
\end{equation}
where $|Dm\rangle$ and $|Nm\rangle$ denote degenerate and non-degenerate states
respectively.  Operators $\hat L^{cl}_{x, y}$ together with
$\hat L^{cl}_z=(1-\hat S_{\bf n}^2)\hat K_z(1-\hat S_{\bf n}^2)$ form
an $su(2)$
algebra
\begin{equation}
[\hat L^{cl}_i, \hat L^{cl}_j]=i\varepsilon_{ijk}\hat L^{cl}_k+O(1/K)
\label{su2cl}
\end{equation}
The degenerate subspace transforms under a $2K+1$ dimensional representation
of this $su(2)$, while nondegenerate states are singlets.
Neglecting terms of the order of $1/K$, we can write Hamiltonian
(\ref{Hcl}) as
\begin{equation}
\hat H_{cl}=\bg\hat S_{\bf n}
\label{HGzcl}
\end{equation}
Note that the operator $\bg$ defined by equation (\ref{bg})
can be interpreted as an ``effective magnetic field operator'', which
explains
 notation (\ref{bg}).
Combining equation (\ref{HGzcl}) with  spin-1 identity (\ref{spiden}),
we obtain
\begin{equation}
\hat S_{\bf n}^2=\frac{\hat H_{cl}^2}{\bg^2}\quad
\hat P_{cl}=1-\frac{\hat H_{cl}^2}{\bg^2}
\label{szclass}
\end{equation}
\begin{equation}
\hat H_{cl}\left[\bg^2-\hat H_{cl}^2\right]=0
\label{idclass}
\end{equation}

Now let us compare these results to the ``quantum'' case. First, one
can check that operators $\hat L^{cl}_i$, $\hat H_{cl}$, and $\hat P_{cl}$
defined above
are indeed the limits of the  corresponding operators introduced in
the previous section. Namely,
\begin{equation}
\hat H\to\hat H_{cl}\quad \hat P\to\hat P_{cl}\quad
\hat J_i\to \hat N_i\to\hat L_i\to  \hat L^{cl}_i\quad\mbox{as $K\to\infty$}
\label{limit}
\end{equation}
Therefore, we can think of the operator
\begin{equation}
\hat I_n^2\equiv \frac{(\hat H+1/2)(\hat H+1)}{\bg^2},
\label{In}
\end{equation}
appearing in the definition of $\hat P$, equation (\ref{Pn}), as a
projection of a spin one onto a ``quantum axis'' along $(K+1/2){\bf
z}+{\bf\hat K}$. Further, the spin-1 identity, equation (\ref{spiden}), and its
descendent equation (\ref{idclass}) are the limit of equation
(\ref{idfactored}).
Equation (\ref{limit}) shows that
 the difference between $\hat N_i$ and $\hat J_i$
disappears in the limit of large $K$. Thus, the gap in the values
of $m$ has no analogs in the leading order of this limit. Note
also that in the leading order in $1/K$ the middle levels of
Hamiltonian (\ref{Hx}) are degenerate at any $x$ and for all
integer values of $S$. An analysis of the subleading corrections
to the large $K$ limit may provide a simple explanation
of why the degeneracies occur only for $S=1$.

\section{Generalized Symmetries}

In this section we deviate somewhat from the main topic of the
paper and discuss  generalized symmetries  of Hamiltonian (\ref{Hx}).
Generalized symmetries are anticommutation relations  that
 explain,  for example, the
following relation between energies of top and bottom levels (see
Section 2):
$$
E_{Tm}+E_{Bm}=-3/2\qquad |m|<K-1
$$
We also show that the
analogy to the large $K$ limit can be pushed even further if we replace the
Hamiltonian (\ref{Hx}) at $x=1$ with an effective Hamiltonian by
keeping only the blocks with $|m|<K-1$.

First, we note that there
exists a basis where the block of the Hamiltonian (\ref{block}) at
$x=1$  reads
\begin{equation}
\left[\begin{array}{ccc}
-1 &\bm&0\\
\bm& -1/2&0\\
0&  0 & -1/2\\
\end{array}\right]
\label{block1}
\end{equation}
Indeed,  matrix (\ref{block1}) has the same eigenvalue equation as
matrix (\ref{block}) at $x=1$. Therefore, matrices (\ref{block1})
and (\ref{block}) represent the same operator in two different
basises. From equation (\ref{In}) we find that $\hat I_{\bf n}^2$
for $|m|<K$ in this basis is
\begin{equation}
 \hat I_{\bf n}^2=\left[\begin{array}{ccc}
1 & 0 & 0\\
0&1&0\\
0&0&0\\
\end{array}\right]
\label{Iz2n}
\end{equation}
Comparing matrix (\ref{block1}) and equation (\ref{HGzcl}), and
using (\ref{Iz2n}), we
conclude that it is natural to define $\hat I_{\bf n}$ as
\begin{equation}
\hat I_{\bf n}=\left[\begin{array}{ccc}
0 & 1 & 0\\
1&0&0\\
0&0&0\\
\end{array}\right]
\label{I1}
\end{equation}

Thus, we can
write (cf. (\ref{HGzcl}))
\begin{equation}
\hat H_{eff}=\bg\hat I_1-1/4\hat I_3-1/4
\hat I_1^2-1/2
\label{su3}
\end{equation}
where $\hat I_1$, $\hat I_2$, and $\hat I_3$
are the first three Gell-Mann matrices of $su(3)$.
$$
\hat I_1\equiv \hat I_{\bf n}=\left[\begin{array}{ccc}
0 & 1 & 0\\
1&0&0\\
0&0&0\\
\end{array}\right]\quad
\hat I_2=\left[\begin{array}{ccc}
0 & -i & 0\\
i&0&0\\
0&0&0\\
\end{array}\right]\quad
\hat I_3=\left[\begin{array}{ccc}
1 & 0 & 0\\
0&-1&0\\
0&0&0\\
\end{array}\right]
$$
Matrices $\hat I_1$, $\hat I_2$, and $\hat I_3$ generate an $su(2)$
subgroup of $su(3)$
known as the isospin subgroup. We note that
\begin{equation}
\hat H_{eff}\hat I_2 +\hat I_2\hat H_{eff}=-\frac{3}{2}\hat I_2
\label{anti}
\end{equation}
Anticommutation relations of this type are sometimes called generalized
symmetries. Equation (\ref{anti}) shows that if $|m\rangle$ is an eigenstate of
the Hamiltonian at $x=1$ with $|m|<K$ and  the energy $E$, $\hat I_2|m\rangle$
is either also an eigenstate but
with the energy $-3/2-E$ or $\hat I_2|m\rangle=0$.
 Examining the eigenstates of matrix (\ref{block1}) one can  verify
that $\hat I_2|Nm\rangle\ne0$ and
$\hat I_2|Dm\rangle=0$. Therefore, we get
\begin{equation}
E_{Tm}+E_{Bm}=-\frac{3}{2}\quad \hat I_2|Tm\rangle=|Bm\rangle\quad
\hat I_2|Bm\rangle=|Tm\rangle\quad \hat I_2|Dm\rangle=0
\label{gen}
\end{equation}
Note also that since for each $|m|<K$ the operator $\k\cdot\s$ has
three eigenvalues $-K-1$, $-1$, and $K$
\begin{equation}
E_{Tm}+E_{Dm}+E_{Bm}=Tr(\hat H)=(K+1/2) Tr(\hat S_z)+Tr(\k\cdot\s)=-2
\label{sum}
\end{equation}

\section{Hydrogen Atom}

Here we derive the nontrivial conservation laws for the Hydrogen atom from
the general expression~(\ref{main}).

The Hamiltonian in atomic units
(see e.g. \cite{LL}) is
\begin{equation}
\hat H=\frac{{\bf\hat p}^2}{2}-\frac{1}{\hat r}\label{H2}
\end{equation}
In addition to the angular momentum ${\bf \hat L}={\bf\hat
  r}\times{\bf\hat p}$, Hamiltonian (\ref{H2}) also conserves the
 Runge--Lenz vector
\begin{equation}
{\bf\hat A}=\frac{\bf\hat r}{r}-\frac{1}{2}({\bf\hat p}\times{\bf \hat
  L}-
{\bf\hat  L}\times{\bf\hat p})
\label{runge}
\end{equation}
Integrals of motion (\ref{runge}) explain the degeneracy in angular
momentum. Components of  vectors ${\bf\hat L}$
and ${\bf\hat A}$ can be combined to yield
 generators of the $so(4)$ symmetry \cite{Pauli} of Hamiltonian (\ref{H2}).

Let us show that  conservation laws (\ref{runge}) can be derived
directly from equation (\ref{main}) using only the knowledge of exact
eigenstates of Hamiltonian (\ref{H2}).
The conservation of angular momentum is not specific to $1/r$
potential. We therefore assume that we know that the angular momentum
is conserved, but do not know any other conservation laws of
Hamiltonian (\ref{H2}).

According to (\ref{main}) the conservation laws are of the form
\begin{equation}
\hat X=\sum_{n} \hat P_n\hat Y\hat P_n\label{main1}
\end{equation}
where $\hat P_n$ projects out the eigenspace of (\ref{H2}) with eigenvalue
$E_n=-1/2n^2$. We take a straightforward approach and try the simplest
choices for $\hat Y$.

First, we look for a conserved scalar and try $\hat Y_1={\bf\hat r}^2$ and
$\hat Y_2={\bf\hat p}^2$. This choice as well as any other powers of
$|{\bf\hat p}|$ and $|{\bf\hat r}|$ result only in trivial commuting
operators - combinations of $\hat H$ and ${\bf\hat L^2}$. For example,
$$
\hat X_1=\sum_n \hat P_n {\bf\hat r}^2 \hat P_n=
\frac{1}{4\hat H}\left(\frac{5}{2\hat H}+
{\bf \hat L}^2-1\right)
$$

Next, we search for a conserved vector and try $\hat Y={\bf\hat r}$. To
evaluate
\begin{equation}
{\bf\hat B}=\sum_n \hat P_n{\bf\hat r} \hat P_n\label{rungeb}
\end{equation}
 explicitly it is advantageous to use parabolic coordinates. The
 eigenstates of discrete spectrum  are specified by
 three integers $|n_1, n_2, m\rangle$, where $m$ is the z-projection of
 the angular momentum and the principle quantum number is
$n=n_1+n_2+|m|+1$ (see e.g. \cite{LL}). Evaluating the matrix
elements of $B_z$  in this basis, we find that only diagonal
matrix elements are nonzero and
\begin{equation}
\hat B_z|n_1, n_2, m\rangle=\frac{3n}{2}(n_1-n_2)|n_1, n_2, m\rangle
\label{Bz}
\end{equation}
To express $\hat B_z$ through components of ${\bf\hat r}$ and
${\bf\hat p}$
we have
to identify operators that have eigenvalues $n$, $n_1$, and $n_2$. All we need
to do  is to follow backwards the derivation of eigenstates in
parabolic coordinates. Substituting these operators instead of $n$,
$n_1$, and $n_2$ into equation (\ref{Bz}), we find
$$
\hat B_z=-\frac{3\hat A_z}{4\hat H}
$$
Since the choice of z-axis is arbitrary, we conclude that
$$
{\bf\hat B}=-\frac{\bf\hat A}{4\hat H}
$$
with ${\bf\hat A}$ given by equation (\ref{runge}).

Thus, the substitution of $\hat Y={\bf\hat r}$
into equation (\ref{main}) produces the Runge-Lenz vector - the
nontrivial conservation law responsible for the ``accidental''
degeneracy in Hydrogen atom. One can check that the alternative choice
$\hat Y={\bf \hat p}$ yields the same result.

\section{Conclusion}

We have derived a complete set of symmetry generators of
Hamiltonian (\ref{Hx}) at $x=1$ and established that the symmetry
group of the Hamiltonian is $SU(2)$. The degenerate
subspace  transforms under a $(2K+1)$-dimensional representation of
this $SU(2)$,  nondegenerate states being singlets. For $K\gg1$
the degenerate states correspond to
$\hat S_{\bf n}=0$ where ${\bf n}$ is a unit vector
along ${\bf z}+\k/K$ and the symmetry operators have a meaning of
rotations of $\k$ for these states. We outlined a  general
approach for finding symmetries that can be used when the
eigenstates and eigenvalues of the Hamiltonian are known. The
formal starting point in this approach is the general expression
for conservation laws (\ref{main}). We have seen that for both
Hamiltonian (\ref{Hx}) and the Hydrogen atom simplest choices for $\hat Y$
produce all nontrivial conservation laws.

We do not have a satisfactory explanation of why multiple degeneracies
 at nonzero value of $x$ occur only for $S=1$ and not for other
 integer values of $S$. If this question is to be answered on the symmetry
 grounds, further insight into the nature and simple manifestations of
symmetries derived in this paper is needed. In particular, the
analysis of subleading corrections to large $K$ limit may be an
interesting avenue to pursue.

\section{Appendix: Commuting Operators}
\setcounter{equation}{0}
\renewcommand{\theequation}{A.\arabic{equation}}

Here we prove  equation (\ref{main}) for Hamiltonian (\ref{Hx}).
The proof can be readily generalized to any Hermitian
operator whose eigenstates span the Hilbert space.

In the energy representation  Hamiltonian (\ref{Hx}) can be written as
\begin{equation}
\hat H=\left[\begin{array}{cc}
 -\frac{1}{2}\cal{E} &  0\\
&\\
  0 &   \cal{D}\\
\end{array}\right]
\label{Hmatrix}
\end{equation}
where $\cal{E}$ is a $(2K+1)\times(2K+1)$ identity matrix and $\cal{D}$ is a
diagonal
matrix with no two diagonal elements equal to each other or to  $-1/2$.
The block $-1/2\cal{E}$
corresponds to
the degenerate subspace, $\cal{D}$ to the remaining levels.
Since any matrix
commutes with an identity matrix  and only diagonal matrices
commute with $\cal{D}$, an operator $\hat X$
commutes
with the Hamiltonian if and only if it is of the form
\begin{equation}
\hat X=\left[\begin{array}{cc}
   \cal{A} &  0 \\
&\\
  0 &   \cal{D}' \\
\end{array}\right]=
\left[\begin{array}{cc}
    \cal{A} &  0 \\
&\\
  0 &   0\\
\end{array}\right]
+
\left[\begin{array}{cc}
   0 &  0 \\
&\\
  0 &   \cal{D}'\\
\end{array}\right]
\label{comgenmat}
\end{equation}
where $\cal{A}$ is an arbitrary $(2K+1)\times(2K+1)$ matrix and $\cal{D}'$ is
a diagonal
matrix of an appropriate size.
Using  the projection operator $\hat P$, we  can rewrite (\ref{comgenmat})  as
\begin{equation}
\hat X=\hat P\hat Y\hat P+\hat\Lambda
\label{comgenop}
\end{equation}
where $\hat Y$ is an arbitrary operator and $\hat\Lambda$ represents
 the second operator on
the left hand side of equation (\ref{comgenmat}). The action of
the operator $\hat\Lambda$ on the degenerate subspace is
trivial. Namely,
$$
\hat\Lambda|Dm\rangle=0
$$
 Thus,
with no loss of generality the search of commuting operators can be
restricted to  operators of the form
\begin{equation}
\hat X=\hat P\hat Y\hat P
\label{commain1}
\end{equation}

\section{Acknowledgment}

We are  grateful to Natalia Saulina  for helpful discussions. B.
L. A. also acknowledges travelling support of the EPSRC under the
grant GR/R33311.

\end{document}